\documentclass[12pt]{article}
\usepackage{epsfig}
\usepackage{amsmath}
\usepackage{hhline}
\usepackage{amssymb}
\usepackage{times}
\usepackage{cite}

\newlength{\dinwidth}
\newlength{\dinmargin}
\begin{document}  
\newcommand{\pom}{{I\!\!P}}
\newcommand{\reg}{{I\!\!R}}
\newcommand{\slowpi}{\pi_{\mathit{slow}}}
\newcommand{\fiidiii}{F_2^{D(3)}}
\newcommand{\fiidiiiarg}{\fiidiii\,(\beta,\,Q^2,\,x)}
\newcommand{\n}{1.19\pm 0.06 (stat.) \pm0.07 (syst.)}
\newcommand{\nz}{1.30\pm 0.08 (stat.)^{+0.08}_{-0.14} (syst.)}
\newcommand{\fiidiiiful}{F_2^{D(4)}\,(\beta,\,Q^2,\,x,\,t)}
\newcommand{\fiipom}{\tilde F_2^D}
\newcommand{\ALPHA}{1.10\pm0.03 (stat.) \pm0.04 (syst.)}
\newcommand{\ALPHAZ}{1.15\pm0.04 (stat.)^{+0.04}_{-0.07} (syst.)}
\newcommand{\fiipomarg}{\fiipom\,(\beta,\,Q^2)}
\newcommand{\pomflux}{f_{\pom / p}}
\newcommand{\nxpom}{1.19\pm 0.06 (stat.) \pm0.07 (syst.)}
\newcommand {\gapprox}
   {\raisebox{-0.7ex}{$\stackrel {\textstyle>}{\sim}$}}
\newcommand {\lapprox}
   {\raisebox{-0.7ex}{$\stackrel {\textstyle<}{\sim}$}}
\def\gsim{\,\lower.25ex\hbox{$\scriptstyle\sim$}\kern-1.30ex%
\raise 0.55ex\hbox{$\scriptstyle >$}\,}
\def\lsim{\,\lower.25ex\hbox{$\scriptstyle\sim$}\kern-1.30ex%
\raise 0.55ex\hbox{$\scriptstyle <$}\,}
\newcommand{\pomfluxarg}{f_{\pom / p}\,(x_\pom)}
\newcommand{\dsf}{\mbox{$F_2^{D(3)}$}}
\newcommand{\dsfva}{\mbox{$F_2^{D(3)}(\beta,Q^2,x_{I\!\!P})$}}
\newcommand{\dsfvb}{\mbox{$F_2^{D(3)}(\beta,Q^2,x)$}}
\newcommand{\dsfpom}{$F_2^{I\!\!P}$}
\newcommand{\gap}{\stackrel{>}{\sim}}
\newcommand{\lap}{\stackrel{<}{\sim}}
\newcommand{\fem}{$F_2^{em}$}
\newcommand{\tsnmp}{$\tilde{\sigma}_{NC}(e^{\mp})$}
\newcommand{\tsnm}{$\tilde{\sigma}_{NC}(e^-)$}
\newcommand{\tsnp}{$\tilde{\sigma}_{NC}(e^+)$}
\newcommand{\st}{$\star$}
\newcommand{\sst}{$\star \star$}
\newcommand{\ssst}{$\star \star \star$}
\newcommand{\sssst}{$\star \star \star \star$}
\newcommand{\tw}{\theta_W}
\newcommand{\sw}{\sin{\theta_W}}
\newcommand{\cw}{\cos{\theta_W}}
\newcommand{\sww}{\sin^2{\theta_W}}
\newcommand{\cww}{\cos^2{\theta_W}}
\newcommand{\trm}{m_{\perp}}
\newcommand{\trp}{p_{\perp}}
\newcommand{\trmm}{m_{\perp}^2}
\newcommand{\trpp}{p_{\perp}^2}
\newcommand{\alp}{\alpha_s}

\newcommand{\alps}{\alpha_s}
\newcommand{\sqrts}{$\sqrt{s}$}
\newcommand{\LO}{$O(\alpha_s^0)$}
\newcommand{\Oa}{$O(\alpha_s)$}
\newcommand{\Oaa}{$O(\alpha_s^2)$}
\newcommand{\PT}{p_{\perp}}
\newcommand{\JPSI}{J/\psi}
\newcommand{\sh}{\hat{s}}
\newcommand{\uh}{\hat{u}}
\newcommand{\MP}{m_{J/\psi}}
\newcommand{\PO}{I\!\!P}
\newcommand{\xbj}{x}
\newcommand{\xpom}{x_{\PO}}
\newcommand{\ttbs}{\char'134}
\newcommand{\xpomlo}{3\times10^{-4}}  
\newcommand{\xpomup}{0.05}  
\newcommand{\dgr}{^\circ}
\newcommand{\pbarnt}{\,\mbox{{\rm pb$^{-1}$}}}
\newcommand{\gev}{\,\mbox{GeV}}
\newcommand{\WBoson}{\mbox{$W$}}
\newcommand{\fbarn}{\,\mbox{{\rm fb}}}
\newcommand{\fbarnt}{\,\mbox{{\rm fb$^{-1}$}}}
\newcommand{\dsdx}[1]{$d\sigma\!/\!d #1\,$}
\newcommand{\eV}{\mbox{e\hspace{-0.08em}V}}
%
%
\newcommand{\qsq}{\ensuremath{Q^2} }
\newcommand{\gevsq}{\ensuremath{\mathrm{GeV}^2} }
\newcommand{\et}{\ensuremath{E_t^*} }
\newcommand{\rap}{\ensuremath{\eta^*} }
\newcommand{\gp}{\ensuremath{\gamma^*}p }
\newcommand{\dsiget}{\ensuremath{{\rm d}\sigma_{ep}/{\rm d}E_t^*} }
\newcommand{\dsigrap}{\ensuremath{{\rm d}\sigma_{ep}/{\rm d}\eta^*} }

\newcommand{\dstar}{\ensuremath{D^*}}
\newcommand{\dstarp}{\ensuremath{D^{*+}}}
\newcommand{\dstarm}{\ensuremath{D^{*-}}}
\newcommand{\dstarpm}{\ensuremath{D^{*\pm}}}
\newcommand{\zDs}{\ensuremath{z(\dstar )}}
\newcommand{\Wgp}{\ensuremath{W_{\gamma p}}}
\newcommand{\ptds}{\ensuremath{p_t(\dstar )}}
\newcommand{\etads}{\ensuremath{\eta(\dstar )}}
\newcommand{\ptj}{\ensuremath{p_t(\mbox{jet})}}
\newcommand{\ptjn}[1]{\ensuremath{p_t(\mbox{jet$_{#1}$})}}
\newcommand{\etaj}{\ensuremath{\eta(\mbox{jet})}}
\newcommand{\detadsj}{\ensuremath{\eta(\dstar )\, \mbox{-}\, \etaj}}

\def\Journal#1#2#3#4{{#1} {\bf #2} (#3) #4}
\def\NCA{\em Nuovo Cimento}
\def\NIM{\em Nucl. Instrum. Methods}
\def\NIMA{{\em Nucl. Instrum. Methods} {\bf A}}
\def\NPB{{\em Nucl. Phys.}   {\bf B}}
\def\PLB{{\em Phys. Lett.}   {\bf B}}
\def\PRL{\em Phys. Rev. Lett.}
\def\PRD{{\em Phys. Rev.}    {\bf D}}
\def\ZPC{{\em Z. Phys.}      {\bf C}}
\def\EJC{{\em Eur. Phys. J.} {\bf C}}
\def\CPC{\em Comp. Phys. Commun.}

\newcommand{\be}{\begin{equation}} 
\newcommand{\ee}{\end{equation}} 
\newcommand{\ba}{\begin{eqnarray}} 
\newcommand{\ea}{\end{eqnarray}}

\begin{titlepage}

\noindent
\begin{flushleft}
{\tt DESY 08-052    \hfill    ISSN 0418-9833} \\
{\tt May 2008}                  \\
\end{flushleft}

\vspace{2cm}
\begin{center}
\begin{Large}

{\bf Search for Excited Electrons in \begin{boldmath}$ep$\end{boldmath} Collisions at HERA \\}

\vspace{2cm}

H1 Collaboration

\end{Large}
\end{center}

\vspace{2cm}

\begin{abstract}
A search for excited electrons is performed using the full  $e^{\pm}p$~data sample collected by the H1 experiment at HERA, corresponding to a total luminosity of $475$~pb$^{-1}$.
The electroweak decays of excited electrons ${e}^{*}\rightarrow{e}{\gamma}$, ${e}^{*}\rightarrow{e}Z$ and ${e}^{*}{\rightarrow}\nu W$ with subsequent hadronic or leptonic decays of the $W$ and $Z$ bosons are considered.
No evidence for excited electron production is found. 
Mass dependent exclusion limits on $e^*$ production cross sections and on the ratio $f/{\Lambda}$ of the coupling to the compositeness scale are derived within gauge mediated models.
These limits extend the excluded region compared to previous excited electron searches.
The $e^*$ production via contact interactions is also addressed for the first time in $ep$ collisions.
\end{abstract}

\vspace{1.5cm}

\begin{center}
Submitted to \PLB
\end{center}

\end{titlepage}

%
%
%
\begin{flushleft}

F.D.~Aaron$^{5,49}$,           
C.~Alexa$^{5}$,                
V.~Andreev$^{25}$,             
B.~Antunovic$^{11}$,           
S.~Aplin$^{11}$,               
A.~Asmone$^{33}$,              
A.~Astvatsatourov$^{4}$,       
A.~Bacchetta$^{11}$,           
S.~Backovic$^{30}$,            
A.~Baghdasaryan$^{38}$
% ...

\bigskip{\it
 $ ^{1}$ I. Physikalisches Institut der RWTH, Aachen, Germany$^{ a}$ \\
% ...
$ ^{4}$ Inter-University Institute for High Energies ULB-VUB, Brussels;
          Universiteit Antwerpen, Antwerpen; Belgium$^{ c}$ \\
 $ ^{5}$ National Institute for Physics and Nuclear Engineering (NIPNE) ,
          Bucharest, Romania \\
% ...
$ ^{11}$ DESY, Hamburg, Germany \\
% ...
$ ^{25}$ Lebedev Physical Institute, Moscow, Russia$^{ e}$ \\
% ...
$ ^{30}$ Faculty of Science, University of Montenegro,
          Podgorica, Montenegro$^{ e}$ \\
% ...
$ ^{33}$ Dipartimento di Fisica Universit\`a di Roma Tre
          and INFN Roma~3, Roma, Italy \\
% ...
$ ^{38}$ Yerevan Physics Institute, Yerevan, Armenia \\
% ...
$ ^{49}$ Also at Faculty of Physics, University of Bucharest,
          Bucharest, Romania \\
% ...
$ ^a$ Supported by the Bundesministerium f\"ur Bildung und Forschung, FRG,
      under contract numbers 05 H1 1GUA /1, 05 H1 1PAA /1, 05 H1 1PAB /9,
      05 H1 1PEA /6, 05 H1 1VHA /7 and 05 H1 1VHB /5 \\
% ...
$ ^c$ Supported by FNRS-FWO-Vlaanderen, IISN-IIKW and IWT
      and  by Interuniversity
Attraction Poles Programme,
      Belgian Science Policy \\
% ...
$ ^e$ Supported by the Deutsche Forschungsgemeinschaft \\
% ...
}

\end{flushleft}
%

\newpage

\section{Introduction}

The three-family structure and mass hierarchy of the known fermions is one of the most puzzling characteristics of the Standard Model (SM) of particle physics.
Attractive explanations are provided by models assuming composite quarks and leptons~\cite{Harari:1982xy}.
The existence of excited states of leptons and quarks is a natural consequence of these models and their discovery would provide convincing evidence of a new scale of matter.
Electron\footnote{In this letter the term ``electron'' refers to both electron and positrons, if not otherwise stated.}-proton interactions at very high energies provide good conditions to search for excited states of first generation fermions. %
For instance, excited electrons ($e^*$) could be singly produced through the exchange of a $\gamma$ or a $Z$ boson in the  $t$-channel. 

In this letter a search for excited electrons using the complete $e^{\pm}p$  HERA collider data of the H1 experiment is presented.
Electroweak decays into a SM lepton ($e$,  $\nu_e$) and a SM gauge boson ($\gamma$, $W$ and $Z$) are considered and hadronic as well as leptonic decays of the $W$ and $Z$ are analysed.

The data are recorded at electron beam energy of $27.6$~GeV and proton beam energies of $820$~GeV and $920$~GeV, corresponding to centre-of-mass energies $\sqrt{s}$ of $301$~GeV and $319$~GeV, respectively.
The total integrated luminosity of the data is $475$~pb$^{-1}$. 
The data comprise $184$~pb$^{-1}$ recorded in $e^-p$ collisions and $291$~pb$^{-1}$ in $e^+p$ collisions, of which $35$~pb$^{-1}$ were recorded at \mbox{$\sqrt{s} = 301$~GeV}. 
With a four-fold increase in statistics, this analysis supercedes the result of the previous H1 search for excited electrons~\cite{Adloff:2002dy}.
It complements the search for excited neutrinos~\cite{Aaron:2008xe}.

\section{Excited Electron Models}

In the present study a model~\cite{Hagiwara:1985wt,Baur:1989kv,Boudjema:1992em} is considered in which excited fermions are assumed to have spin  $1/2$ and isospin $1/2$.
The left-handed and right-handed components of the excited fermions form weak iso-doublets $F_L^*$ and $F_R^*$.

Interactions between excited and ordinary fermions may be mediated by gauge bosons, as described by the effective Lagrangian~\cite{Baur:1989kv,Boudjema:1992em}:

\be
{\cal L}_{GM} = \frac{1}{2\Lambda}{\bar{F^{*}_{R}}} \; {{\sigma}^{\mu\nu}} \left[ gf\frac{\tau^a}{2}{W_{\mu\nu}^a}+g'f'\frac{Y}{2}B_{\mu\nu} + g_s f_s \frac{\lambda^a}{2} G^a_{\mu\nu} \right]  {F_{L}} + h.c. \; .
\label{eq:lagrangian}
\ee

Only the right-handed component of the excited fermion $F_R^*$ is allowed to couple to light fermions, in order to protect the light leptons from radiatively acquiring a large anomalous magnetic moment~\cite{Brodsky:1980zm,Renard:1982ij}.
The matrix ${{\sigma}^{\mu\nu}}$ is the covariant bilinear tensor, $W_{{\mu}{\nu}}^a$,  $B_{{\mu}{\nu}}$ and $G^a_{\mu\nu}$ are the field-strength tensors of the SU($2$), U($1$) and SU($3$)$_{C}$ gauge fields, $\tau^a$, $Y$ and $\lambda^a$ are the Pauli matrices, the weak hypercharge operator and the Gell-Mann matrices, respectively. The standard electroweak and strong gauge couplings are denoted by $g$, $g'$ and $g_s$, respectively. 
The parameter $\Lambda$ has units of energy and can be regarded as the compositeness scale which reflects the range of the new confinement force. 
The constants $f$, $f'$ and $f_s$ are coupling parameters associated to the three gauge groups and are determined by the yet unknown composite dynamics. 
\par
Following this model of gauge mediated (GM) interactions, single  $e^*$ production in $ep$ collisions may result from the $t$-channel exchange of a $\gamma$ or $Z$ boson. 
Since the $e^*$ is expected not to have strong interactions, the present search is insensitive to $f_s$.
The produced $e^*$  may decay into a lepton and an electroweak gauge boson via $e^* {\rightarrow} e\gamma$, $e^* {\rightarrow} \nu W$ and $e^* {\rightarrow} e Z$.
For a given $e^*$ mass  value $M_{e^*}$ and assuming a numerical relation between $f$ and $f'$, the $e^*$ branching ratios are fixed and the production cross section depends only on $f/\Lambda$.
In most analyses the assumption is made that the coupling parameters  $f$ and $f'$ are of comparable strength and only the relationships $f = - f'$ and $f = + f'$ are considered.
In the case $f = - f'$, the excited electron does not couple to the photon and therefore the $e^*$ production cross section at HERA is small.
Therefore, only the case $f = + f'$ is considered in this analysis.  

In addition to GM interactions, novel composite dynamics may be visible as contact interactions (CI) between excited fermions and SM quarks and leptons.
Such interactions can be described by the effective four-fermion Lagrangian~\cite{Baur:1989kv}:

\be
{\cal L}_{CI} = \frac{4 \, \pi}{2 \, \Lambda^2} \; j^\mu j_\mu \; ,
\label{eq:lagrangian_CI}
\ee

\noindent where $\Lambda$ is assumed to be the same parameter as in the Lagrangian (\ref{eq:lagrangian}) and $j_\mu$ is the fermion current

\be
j_\mu =  \eta_L \bar{F_L^*} \gamma_\mu F_L +  \eta'_L \bar{F_L} \gamma_\mu F_L + \eta''_L \bar{F_L^*} \gamma_\mu F_L^*
+  h.c. + \left( L \rightarrow R \right).
\label{eq:lagrangian_CI2}
\ee

Conventionally, the $\eta$ factors are set to one for the left-handed and to zero for the right-handed current.

Contact interactions may induce changes in the cross section of neutral current (NC) deep-inelastic scattering (DIS) $ep \rightarrow e X$.
Searches for deviations from the SM cross section at the highest squared momentum transfers $Q^2$ in NC DIS processes have excluded values of $\Lambda$ between $1.6$~TeV and $5.5$~TeV, depending on the chiral structure considered~\cite{Adloff:2003jm}.
Contact interactions may also mediate the resonant production of excited electrons in $ep$ collisions as well as their decays into an electron and a pair of SM fermions.
The $e^*$ production and decay by both gauge and contact interactions is also considered in this analysis.
In this case the total $e^*$ production cross section $\sigma_{GM+CI}$ is the sum of pure GM and CI cross sections and of the interference between the two processes~\cite{Spira}.
For simplicity, the relative strength of gauge and contact interactions are fixed by setting the parameters $f$ and $f'$ of the gauge interaction to one. 
The ratio of the GM+CI and GM cross sections $\sigma_{GM+CI}/\sigma_{GM}$ then depends only on $\Lambda$ and on the $e^*$ mass. For $M_{e^*} = 150$~GeV and $\Lambda = 1$~TeV,  $\sigma_{GM+CI}/\sigma_{GM}$ is equal to $8.4$, but reduces to  $1.3$ for $\Lambda = 4$~TeV.
Relative branching ratios of GM and CI decays are determined by the $e^*$ partial widths in each decay channel~\cite{Baur:1989kv}.
In the sensitive domain of the present analysis  ($\Lambda \simeq 4$~TeV and $100$~GeV~$< M_{e^*} <$~$200$~GeV), more than $95$\% of $e^*$ decays are gauge mediated. 
Therefore, only GM decay channels are used for the present search.

\section{Simulation of Signal and Background Processes}

The Monte Carlo (MC) event generator COMPOS~\cite{Kohler:1991yu} is used for the calculation of the $e^*$ production cross section and to determine the signal detection efficiencies.
It is based on the cross section formulae for gauge mediated interactions~\cite{Hagiwara:1985wt,Baur:1989kv}.
Cross section formulae for contact interaction production and for the interference between contact and gauge interactions~\cite{Spira} have also been incorporated into COMPOS. 
Only $e^*$ decays via gauge mediated interactions are simulated. 
Initial state radiation of a photon from the incident electron is included using the Weizs\"acker--Williams approximation~\cite{Berger:1986ii}. 
The proton parton densities are taken from the CTEQ5L~\cite{Pumplin:2002vw} parametrisation and are evaluated at the scale $\sqrt{Q^2}$.
The parton shower approach~\cite{Sjostrand:2000wi} is applied in order to simulate Quantum Chromodynamics (QCD) corrections in the initial and final states. Hadronisation is performed using Lund string fragmentation as implemented in PYTHIA~\cite{Sjostrand:2000wi}.
The COMPOS generator uses the narrow width approximation (NWA) for the calculation of the production cross section and takes into account the natural width of the excited electron for the $e^*$ decay.
The NWA is valid for $e^{*}$ masses below $290$~GeV and the couplings $f/\Lambda$ relevant to this analysis, as the total $e^{*}$ width is less than $10$\% of the $e^*$ mass.

The Standard Model (SM) processes which may mimic the $e^*$ signal are QED Compton scattering, neutral current and charged current (CC) deep-inelastic scattering and to a lesser extent photoproduction, lepton pair production and real $W$ boson production. 

The RAPGAP~\cite{Jung:1993gf} event generator, which implements the Born, QCD Compton and Boson Gluon Fusion matrix elements, is used to model NC DIS events. 
The QED radiative effects arising from real photon emission from both the incoming and outgoing electrons are simulated using the HERACLES~\cite{Kwiatkowski:1990es} program. 
Direct and resolved photoproduction of jets and prompt photon production are simulated using the PYTHIA event generator. 
The simulation is based on Born level hard scattering matrix elements with radiative QED corrections. 
In RAPGAP and PYTHIA, jet production from higher order QCD radiation is simulated using leading logarithmic parton showers and hadronisation is modelled with Lund string fragmentation.
The leading order MC prediction of NC DIS and photoproduction processes with two or more high transverse momentum jets is scaled by a factor of $1.2$ to account for missing higher order QCD contributions in the MC generators~\cite{Adloff:2002au,Aktas:2004pz}. 
Charged current DIS events are simulated using the DJANGO~\cite{Schuler:yg} program, which includes first order leptonic QED radiative corrections based on HERACLES. The production of two or more jets in DJANGO is accounted for using the colour-dipole-model~\cite{Lonnblad:1992tz}. 
Contributions from elastic and quasi-elastic QED Compton scattering are simulated with the WABGEN~\cite{Berger:kp} generator. 
Contributions arising from the production of $W$ bosons and multi-lepton events are modelled using the EPVEC~\cite{Baur:1991pp} and GRAPE~\cite{Abe:2000cv} event generators, respectively.

Generated events are passed through the full GEANT~\cite{Brun:1987ma} based simulation of the H1 apparatus, which takes into account the actual running conditions of the data taking, and are reconstructed and analysed using the same program chain as for the data.

\section{Experimental Conditions}

A detailed description of the H1 experiment can be found in \cite{Abt:h1_1}.
Only the detector components relevant to the
present analysis are briefly described here.  
The origin of the H1 coordinate system is the nominal $ep$ interaction point, with the direction of the proton beam defining the positive $z$-axis (forward region). Transverse momentum ($P_T$) is measured in the $xy$ plane. The pseudorapidity $\eta$ is related to the polar angle $\theta$ by $\eta = -\ln \, \tan (\theta/2)$.
The Liquid Argon (LAr) calorimeter~\cite{Andrieu:1993kh} is used to measure electrons, photons and hadrons. It covers the polar angle range
$4^\circ < \theta < 154^\circ$ with full azimuthal acceptance.
Electromagnetic shower energies are measured with a precision of
$\sigma (E)/E = 12\%/ \sqrt{E/\mbox{GeV}} \oplus 1\%$ and hadronic energies
with $\sigma (E)/E = 50\%/\sqrt{E/\mbox{GeV}} \oplus 2\%$, as measured in test beams~\cite{Andrieu:1994yn,Andrieu:1993tz}.
In the backward region, energy measurements are provided by a lead/scintillating-fiber (SpaCal) calorimeter~\cite{Appuhn:1996na} covering the angular range $155^\circ < \theta < 178^\circ$.
The central ($20^\circ < \theta < 160^\circ$) and forward ($7^\circ < \theta < 25^\circ$)  tracking detectors are used to
measure charged particle trajectories, to reconstruct the interaction
vertex and to complement the measurement of hadronic energy.
The LAr and inner tracking detectors are enclosed in a super-conducting magnetic
coil with a field strength of $1.16$~T.
The return yoke of the coil is the outermost part of the detector and is
equipped with streamer tubes forming the central muon detector
($4^\circ < \theta < 171^\circ$).
In the forward region of the detector ($3^\circ < \theta < 17^\circ$) a set of
drift chambers detects muons and measures their momenta using an iron toroidal magnet.
The luminosity is determined from the rate of the Bethe-Heitler process $ep {\rightarrow} ep \gamma$,
measured using a photon detector located close to the beam pipe at $z=-103~{\rm m}$, in the backward direction.

\section{Data Analysis}\label{sec:anal}

The triggers employed for collecting the data used in this analysis are based on the detection of electromagnetic deposits or missing transverse energy in the LAr calorimeter~\cite{Adloff:2003uh}.
The trigger efficiency is $\sim$~$90$\% for events with missing transverse energy of $20$~GeV, and increases above $95$\% for missing transverse energy above $30$~GeV.
Events containing an electromagnetic deposit (electron or photon) with an energy greater than $10$~GeV are triggered with an efficiency close to $100$\%.

In order to remove background events induced by cosmic showers and other non-$ep$ sources, the event vertex is required to be reconstructed within $35$~cm in $z$ of the nominal interaction point. In addition, topological filters and timing vetoes are applied.

The identification of electrons or photons relies on the measurement of a compact and isolated electromagnetic shower in the LAr calorimeter. 
The hadronic energy within a distance in the pseudorapidity-azimuth $(\eta - \phi)$ plane $R=\sqrt{\Delta \eta^2 + \Delta \phi^2} < 0.5$ around the electron (photon) is required to be below $3$\% of the electron (photon) energy.
Furthermore, each electron (photon) candidate must be isolated from jets by a minimum distance in pseudorapidity-azimuth of $R > 0.5$ to any jet axis.
The electron and photon energy and angular direction are measured by the calorimeters.
Muon identification is based on a track measured in the inner tracking systems associated with signals in the muon detectors~\cite{Andreev:2003pm}.
A muon candidate is required to have no more than $5$~GeV deposited in a
cylinder, centred on the muon track direction, of radius $25$~cm and $50$~cm in the electromagnetic and hadronic sections of the LAr calorimeter, respectively.
Additionally, the muon candidate is required to be separated from the closest jet and from any track by $R > 1$ and  $R > 0.5$, respectively.
Calorimeter energy deposits and tracks not previously identified as electron, photon or muon candidates are used to form combined cluster-track objects, from which the hadronic energy is reconstructed~\cite{matti,benji}.
Jets are reconstructed from these combined cluster-track objects using an inclusive $k_T$ algorithm~\cite{Ellis:1993tq,Catani:1993hr} with a minimum transverse momentum of $2.5$~GeV.
The missing transverse momentum  $P_T^{\rm{miss}}$ of the event is derived from all detected particles and energy deposits in the event.
In events with large $P_T^{\rm{miss}}$, the only non-detected particle in the event is assumed to be a neutrino.  
The four-vector of this neutrino candidate is reconstructed assuming  transverse momentum conservation and the relation $\sum_i (E^i - P_{z}^{i}) + (E^\nu - P_{z}^{\nu}) = 2 E^0_e = 55.2$~GeV, where the sum runs over all detected particles, $P_{z}$ is the momentum along the proton beam axis and $E^0_e$ is the electron beam energy.

Specific selection criteria applied in each decay channel are presented in the following subsections.
A detailed description of the analysis can be found in~\cite{trinh}.

\subsection{\boldmath $e\gamma$ Resonance Search}

The signature of the $e^* {\rightarrow} e \gamma$ decay channel consists of two high $P_T$ isolated electromagnetic clusters. 
SM background arises mainly from elastic and inelastic QED Compton events.
Two isolated electromagnetic clusters are required, each with  transverse momentum $P_T > 15$~GeV and  polar angle $5^\circ < \theta < 130^\circ$.
No explicit electron and photon identification based on tracking conditions is performed in order to retain a high selection efficiency.
To reduce contributions from QED Compton processes, the sum of the energies of the two electromagnetic clusters is required to be greater than $110$~GeV and the sum of their total transverse momenta has to be larger than $75$~GeV.

After this selection, the SM background from elastic QED Compton events is smaller than that from inelastic QED Compton processes. 
Since about half of the $e^*$ production cross section is expected from elastic $e^*$ production~\cite{Hagiwara:1985wt}, the analysis is separated into two parts.
Events with a total hadronic energy $E_h < 5$~GeV are used to search for elastic $e^*$ production, whereas the other events are attributed to possible inelastic $e^*$ production.

\begin{figure}[htbp] 
\includegraphics[width=.5\textwidth]{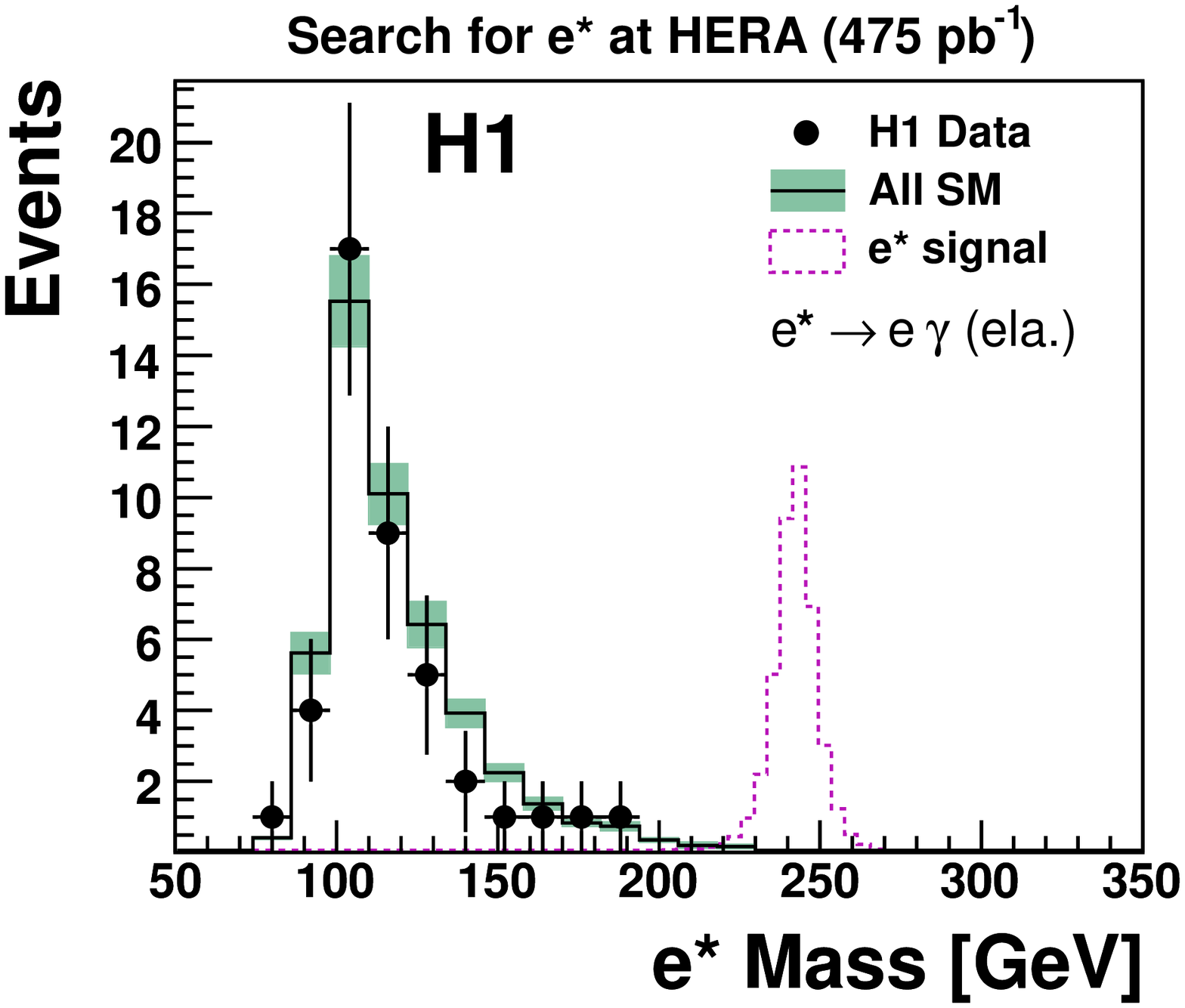}\put(-12,37) {{\bf (a)}}
\includegraphics[width=.5\textwidth]{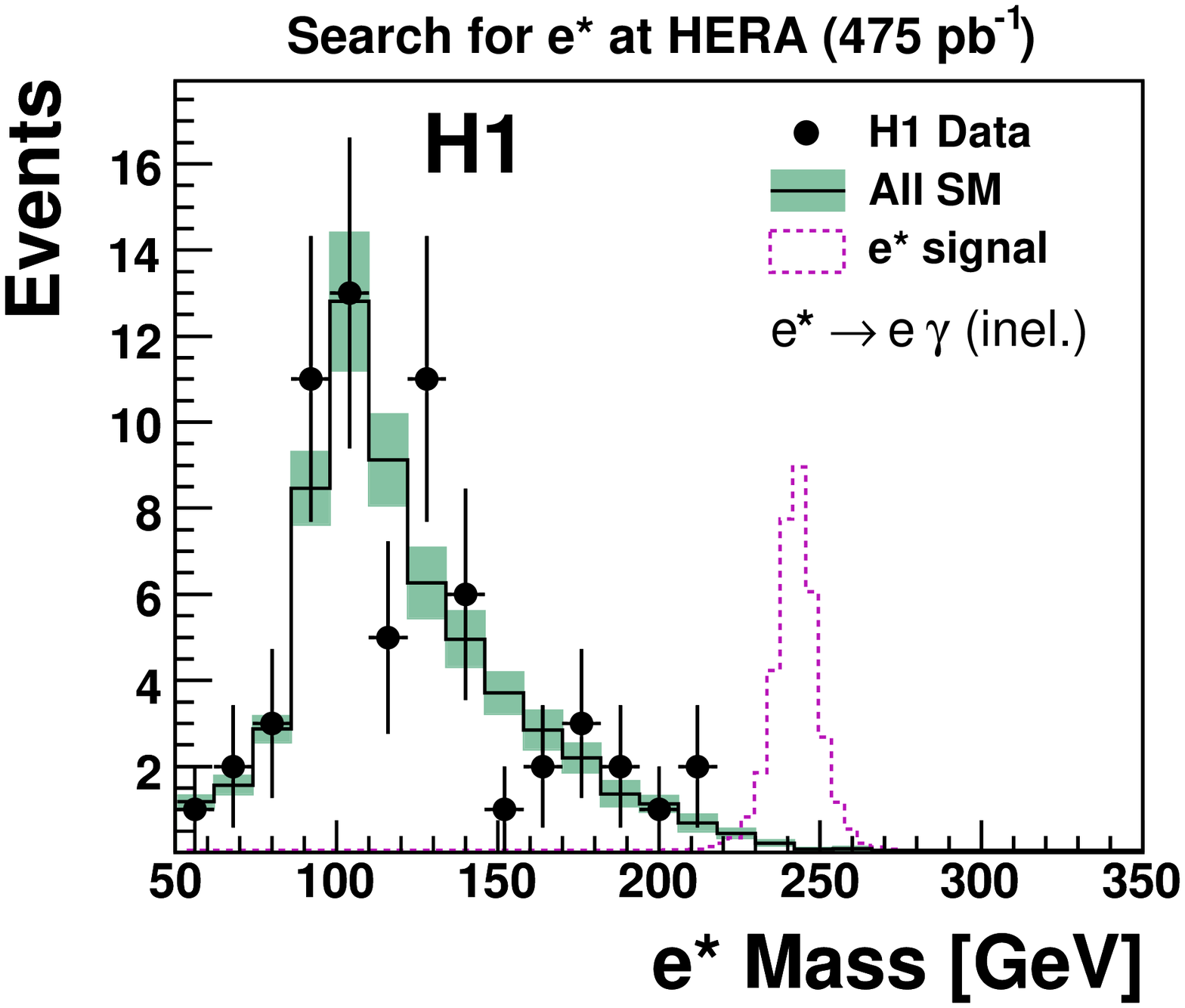}\put(-12,37) {{\bf (b)}}\\
\includegraphics[width=.5\textwidth]{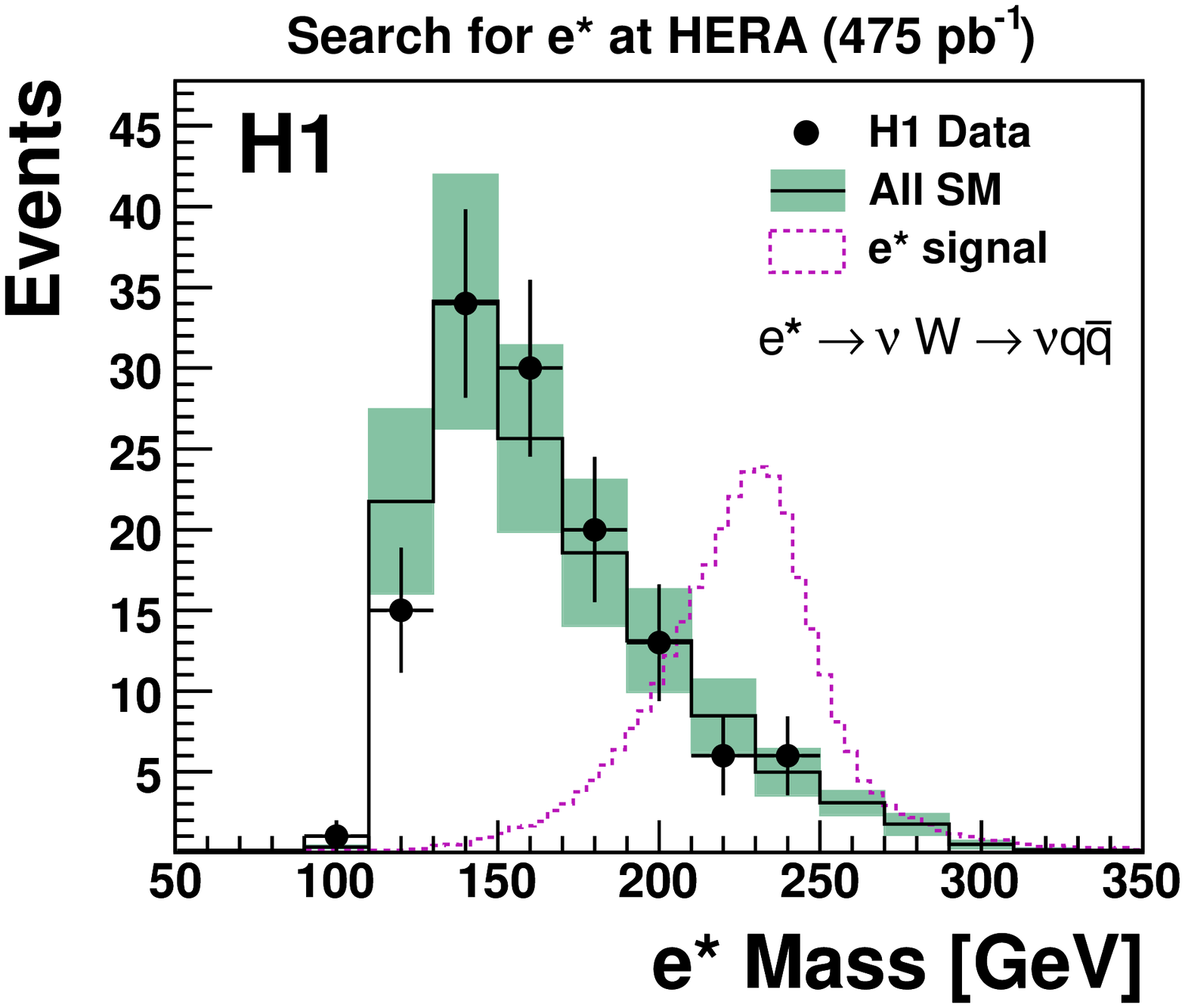}\put(-12,37) {{\bf (c)}} 
\includegraphics[width=.5\textwidth]{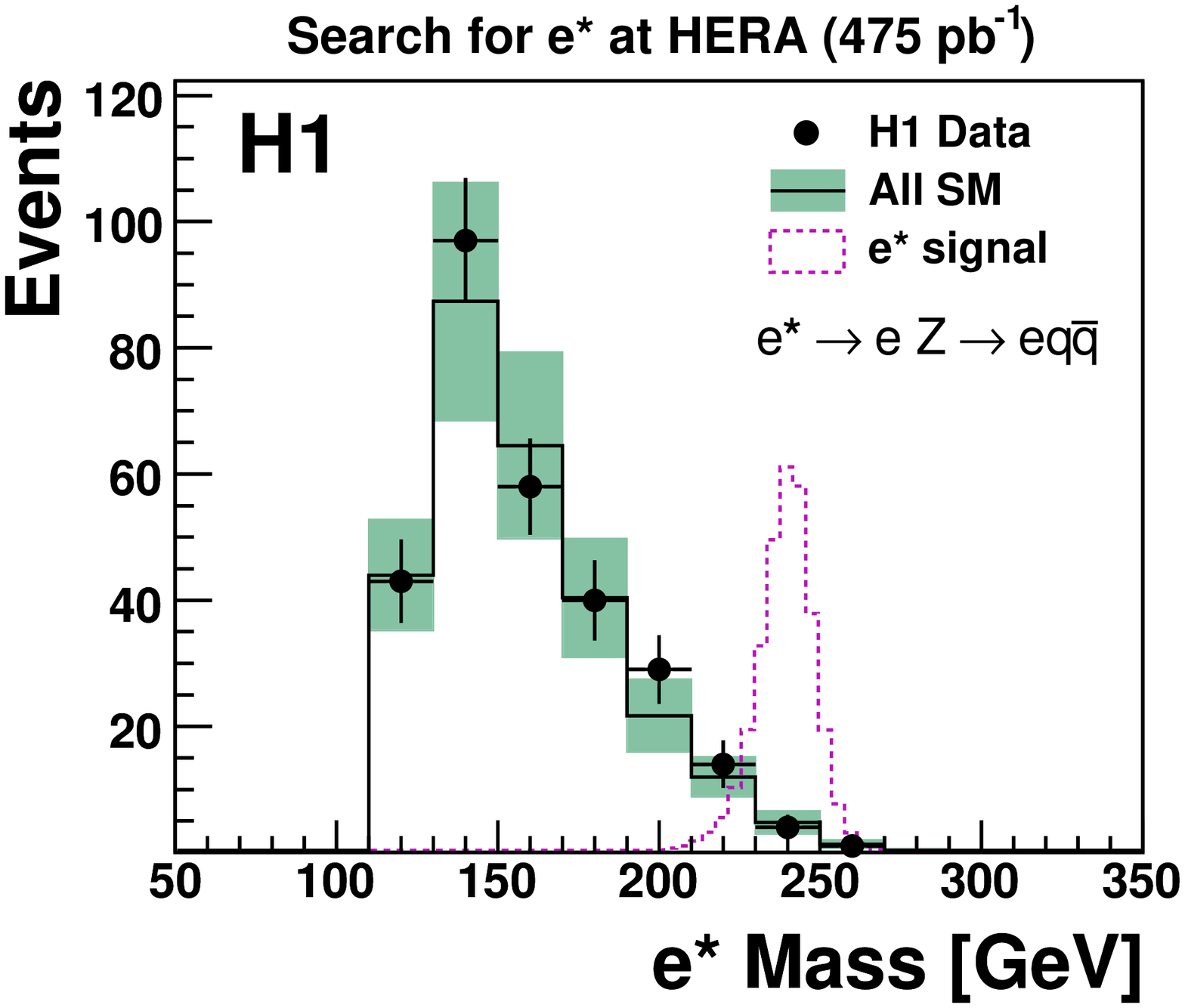}\put(-12,37) {{\bf (d)}}
 \caption{Invariant mass distribution of the $e^*$ candidates in the elastic ${e}^{*} {\rightarrow} {e}{\gamma}$ (a), inelastic ${e}^{*} {\rightarrow} {e}{\gamma}$ (b), ${e}^{*} {\rightarrow} \nu W {\rightarrow} \nu q \bar{q}$ (c), and  ${e}^{*} {\rightarrow} e Z {\rightarrow} e q \bar{q}$ (d) search channels. The points correspond to the observed data events and the histograms to the SM expectation after the final selections. The error bands on the SM prediction include model uncertainties and experimental systematic errors added in quadrature.
The dashed line represents with an arbitrary normalisation the reconstructed mass distribution of $e^*$ events with $M_{e^*} = 240$~GeV.}
 \label{fig:Mass}  
 \end{figure}

In the elastic channel $42$ events are selected in the data compared to a SM expectation of $48$~$\pm$~$4$. In the inelastic channel  $65$ events are found for  $65$~$\pm$~$8$ expected.
The errors on the SM prediction include model and experimental systematic errors added in quadrature (see section \ref{sec:sys_err}).
The invariant mass of the $e^*$ candidate is calculated from the four-vectors of the electron and photon candidates.
The invariant mass distribution of the $e^*$ candidates and the SM background expectations are presented in figure~\ref{fig:Mass}(a) and (b) for the elastic and inelastic channels, respectively.
The selection efficiency is $60$\% for $M_{e^*} = 120$~GeV, increasing to $70$\% for $M_{e^*} = 260$~GeV.
From Monte Carlo studies, the experimental resolution on the reconstructed $e^*$ mass distribution is $3$~GeV for a generated $e^*$ mass of $120$~GeV, increasing to $6$~GeV for an $e^*$ mass of $260$~GeV.

\subsection{\boldmath $\nu{q}{\bar{q}}$  Resonance Search}

The signature of the ${e}^{*} {\rightarrow} \nu W {\rightarrow} \nu q \bar{q}$ decay channel consists of two high transverse momentum jets in events with large $P_T^{\rm{miss}}$.
The SM background is dominated by multi-jet CC DIS events and contains moderate contributions from NC DIS and photoproduction.
Events with missing transverse momentum $P_{T}^{\rm{miss}} > 20$~GeV are selected.
In each event at least two jets with transverse momenta larger than $20$ and $15$~GeV, respectively, are required in the polar angle range $5^{\circ} < \theta < 130^{\circ}$.

The ratio $V_{ap}/V_{p}$ of transverse energy flow anti-parallel and parallel to the hadronic final state~\cite{Adloff:1999ah} is used to suppress photoproduction events. Events with $V_{ap}/V_{p} > 0.3$ are rejected.
Photoproduction and NC DIS backgrounds typically have low  values of $x_h$, the Bjorken scaling variable calculated from the hadronic system using the Jacquet-Blondel method~\cite{Adloff:1999ah,JBmethod}, and are thus suppressed by requiring $x_h > 0.04$.
In each event, a $W$ candidate is reconstructed from the combination of those two jets with invariant mass closest to the nominal $W$ boson mass. 
The reconstructed $W$ candidate is required to have an invariant mass above $60$~GeV. 
In order to further reduce the background from CC DIS, the invariant mass of all jets and hadrons in the event not associated to the decay of the $W$ boson candidate is required to be below $15$~GeV.

After this selection, $129$ events are found compared to a SM expectation of $133$~$\pm$~$32$ events which is dominated by CC DIS events.
The CC DIS cross section is smaller in $e^+p$ collisions than in $e^-p$, in contrast to the $e^*$ cross section which is comparable in both collision modes.
Therefore, $e^+p$ data have a larger sensitivity to a potential $e^*$ signal in this channel than $e^-p$ data.
In the $e^+p$ ($e^-p$) data sample, $33$ ($96$)  events are observed compared to a SM expectation of $51 \pm 13$ ($82 \pm 19$). 
A significant excess is observed neither in $e^+p$ nor in $e^-p$ data.
The invariant mass of the $e^*$ candidate is calculated from the neutrino and $W$ candidate four-vectors.
For this calculation, the $W$ candidate four-vector is scaled such that its mass is set to the nominal $W$ boson mass.
The invariant mass distribution of the $e^*$ candidates and the SM background is presented in figure~\ref{fig:Mass}(c).
The selection efficiency in this channel is  $20$\% for $M_{e^*} = 120$~GeV, increasing to $55$\% for $M_{e^*} = 260$~GeV.
From Monte Carlo studies, the experimental resolution on the reconstructed $e^*$ mass distribution is $9$~GeV for a generated $e^*$ mass of $120$~GeV, increasing to $\sim 20$~GeV for an $e^*$ mass of $260$~GeV.

\subsection{\boldmath $eq{\bar{q}}$ Resonance Search} 

The signature of the ${e}^{*} {\rightarrow} e Z {\rightarrow} e q \bar{q}$ decay channel consists of one electron and two high $P_T$ jets.
Multi-jet NC DIS events constitute the main background contribution from SM processes.
Events are selected with an isolated electron in the LAr calorimeter in the polar angle range $5^\circ< \theta^e < 90^\circ$. 
The electron should have either a transverse momentum $P_T^e$ greater than $25$~GeV or 
the variable\footnote{For NC DIS events, this variable is proportional to the four-momentum transfer squared $Q^2$.} $\xi^e = E^e \cos^2{(\theta^e/2)}$ above $23$~GeV.
These conditions remove a large part of the NC DIS contribution.
The events are required to have at least two jets in the polar angle range $5^{\circ} < \theta^{\rm{jet}} < 130^{\circ}$ with transverse momenta larger than $20$ and $15$~GeV, respectively.
In each event, a $Z$ candidate is reconstructed from the combination of those two jets with invariant mass closest to the nominal $Z$ boson mass. 
The reconstructed mass of the $Z$ candidate is required to be larger than $70$~GeV.
To further reduce the NC DIS background the polar angle of the jet with the highest $P_T$ associated to the $Z$ candidate is required to be less than $80^\circ$.
The polar angle of the second jet is required to be greater than $10^\circ$ in events with $P_T^{\rm{jet}_2} < 25$~GeV.

After this selection, $286$ events are observed while $277$ $\pm$ $62$ are expected from the SM.
The invariant mass of the $e^*$ candidate is calculated from the electron and $Z$ candidate four-vectors.
For this calculation, the $Z$ candidate four-vector is scaled such that its mass is set to the nominal $Z$ boson mass.
The invariant mass distribution of the $e^*$ candidates and the SM background is presented in figure~\ref{fig:Mass}(d).
The selection efficiency in this channel is  $20$\% for $M_{e^*} = 120$~GeV, increasing to $55$\% for $M_{e^*} = 260$~GeV.
From Monte Carlo studies, the experimental resolution on the reconstructed $e^*$ mass distribution is $2$~GeV for a generated $e^*$ mass of $120$~GeV, increasing to $8$~GeV for an $e^*$ mass of $260$~GeV.

\subsection{\boldmath $eee$, $e\mu\mu$ and $e\nu\nu$ Resonance Searches}

In the search for ${e}^{*} {\rightarrow} {e} Z {\rightarrow} eee$, events with three electrons of high transverse momenta are selected.
The electrons must be detected in the polar angle range $5^\circ < \theta^{e} <  150^\circ$ and have transverse momenta larger than $25$, $20$ and $15$~GeV, respectively. 
To reduce the background from QED Compton processes, each electron in the central region ($\theta^e >35^\circ$) must be associated to a charged track.
A $Z$ candidate is reconstructed from the combination of the two electrons with an invariant mass closest to the nominal $Z$ boson mass. 
The reconstructed mass of the $Z$ candidate is required be compatible with the nominal $Z$ boson mass within $7$~GeV.
After this selection no data event remains, while $0.72 \pm 0.06$ SM background events are expected. The selection efficiency for $e^*$ with masses above $120$~GeV is $\sim$ $60$\%.

In the search for ${e}^{*} {\rightarrow} {e} Z {\rightarrow} e \mu \mu$, events are selected with one electron  with transverse momentum above $20$~GeV and two muons with transverse momenta above $15$ and $10$~GeV, respectively.
The electron and the muons must be detected in the polar angle ranges $5^\circ < \theta^{e} <  150^\circ$ and $10^\circ< \theta^{\mu} < 160^\circ$, respectively. 
A $Z$ candidate is reconstructed from the combination of the two muons and its reconstructed mass is required to be larger than $60$~GeV.
After this selection no data event remains, while $0.52 \pm 0.05$ SM background events are expected. 
The selection efficiency in this channel is  $\sim 40$\% for $M_{e^*} = 120$~GeV, decreasing to $15$\% for $M_{e^*} = 260$~GeV.

The signatures of the ${e}^{*} {\rightarrow} {\nu} W {\rightarrow} \nu e \nu$ and ${e}^{*} {\rightarrow} {e} Z {\rightarrow} e \nu \nu$ channels are similar and consist of one high $P_T$ electron in events with large missing transverse momentum.
Events with \mbox{$P_T^{\rm{miss}} > 25$~GeV} and one electron with $P_T > 20$~GeV are selected.
The electron is detected in the polar angle range $5^\circ < \theta^{e} < 100^\circ$ and is required to be isolated from jets by a minimum distance of $R > 1$.
To reduce the background from radiative CC DIS processes, a track must be associated to the electron in the central region ($\theta^e >35^\circ$).
Events from photoproduction are suppressed by requiring $V_{ap}/V_{p} < 0.1$.
Remaining NC DIS events are removed by requiring that the longitudinal momentum balance of the event be $\sum_i (E_i - P_{z,i}) < 45$~GeV, where the sum runs over all visible particles.   
In order to remove background arising from SM $W$ production, the hadron system  is required to have a total transverse momentum $P_T^h < 20$~GeV and to exhibit a polar angle $\gamma_h$, as defined in~\cite{Adloff:1999ah}, below $80^{\circ}$.
In each event, only one neutrino candidate can be reconstructed, from the total missing transverse momentum, as explained at the beginning of section~\ref{sec:anal}. 
The invariant mass of the $e^*$ candidate in the $e \nu \nu$ final state 
is therefore estimated from the four-vectors of the neutrino candidate and the electron candidate. 
To further remove background from $W$ production, only events in which the reconstructed $e^*$ mass is above $90$~GeV are considered.
After this selection four data events remain, while $4.5 \pm 0.7$ SM background events are expected. 
The selection efficiency for the  ${e}^{*} {\rightarrow} {\nu} W {\rightarrow} \nu e \nu$ (${e}^{*} {\rightarrow} {e} Z {\rightarrow} e \nu \nu$) signature is  $\sim 60$\% ($\sim 35$\%) for $e^*$ with masses above $120$~GeV.

\subsection{Systematic Uncertainties}
\label{sec:sys_err}

The following experimental systematic uncertainties are considered:

\begin{itemize}
\item The uncertainty on the electromagnetic energy scale varies between $0.7$\% and $2$\% depending on the polar angle. The polar angle measurement uncertainty is $3$ mrad for electromagnetic clusters.
\item The scale uncertainty on the transverse momentum of high $P_T$ muons amounts to $2.5$\%. The uncertainty on the reconstruction of the muon polar angle is $3$~mrad.
\item The hadronic energy scale is known within $2$\%. The uncertainty on the jet polar angle determination is $10$ mrad.
\item The uncertainty on the trigger efficiency is $3$\%.
\item The luminosity measurement has an uncertainty of $3$\%.
\end{itemize}

The effect of the above systematic uncertainties on the SM expectation and the signal efficiency are determined by varying the experimental quantities by $\pm 1$ standard deviation in the MC samples and propagating these variations through the whole analysis chain.

Additional model systematic uncertainties are attributed to the SM background MC generators described in section $3$.
An error of $20$\% on the normalisation of NC DIS, CC DIS and photoproduction processes with at least two high $P_T$ jets is considered to account for the uncertainty on higher order QCD corrections. 
The error on the elastic and quasi-elastic QED Compton cross sections is conservatively estimated to be $5$\%. 
The error on the inelastic QED Compton cross section is $10$\%.
The errors attributed to lepton pair and $W$ production are $3$\% and $15$\%, respectively.
The total error on the SM background prediction is determined by adding the effects of all model and experimental systematic uncertainties in quadrature.

The theoretical uncertainty on the $e^*$ production cross section is dominated by the uncertainty on the scale at which the proton parton densities are evaluated.
It is estimated by varying this scale from $\sqrt{Q^2}/2$ to $2\sqrt{Q^2}$.
The resulting uncertainty depends on the $e^*$ mass and is $10$\% at $M_{e^*} = 100$~GeV, increasing to $15$\% at $M_{e^*} = 300$~GeV.

\section{Interpretation}

The event yields observed in all decay  channels are in agreement with the corresponding SM expectations and are summarised in table~\ref{tab:estaryields}. %
The SM predictions are dominated by QED Compton for the $e\gamma$ resonance search, by CC DIS  in the $\nu{q}{\bar{q}}$ resonance search and by NC DIS processes for the $e{q}{\bar{q}}$ resonance search. 
The distributions of the invariant mass of the data events are in agreement with those of the expected SM background as shown in figure~\ref{fig:Mass}. 
Few or no data events are observed in channels corresponding to leptonic decays of the $W$ or $Z$ bosons, in agreement with the low SM expectations.

Since no evidence for the production of excited electrons is observed, upper limits on the $e^*$ production cross section and on the model parameters are derived as a function of the mass of the excited electron. 
Limits are presented at the $95$\% confidence level (CL) and are obtained from the mass spectra using a modified frequentist approach which takes statistical and systematic uncertainties into account~\cite{Junk:1999kv}.

\begin{figure}[htbp] 
  \begin{center}
\includegraphics[width=.5\textwidth]{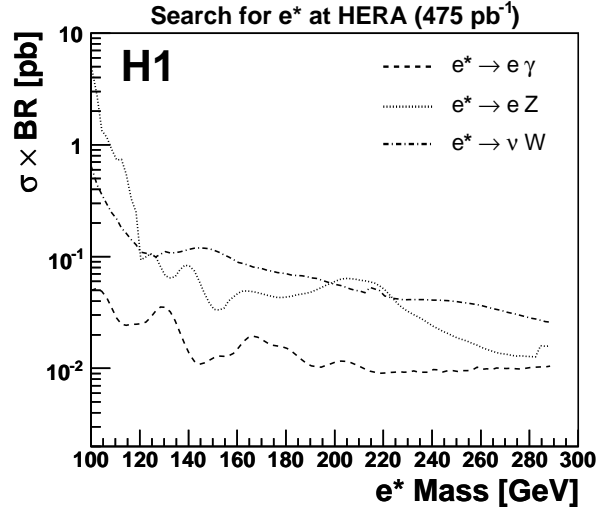}
   \end{center}
  \caption{Upper limits at $95$\% CL on the product of the $e^*$ cross section and decay branching ratio, $\sigma \times$~BR, in the three  $e^*$ decay channels as a function of the excited electron mass. The decay channels of the $W$ and $Z$ gauge bosons are combined. Areas above the curves are excluded. }
\label{fig:LimitSBR}  
\end{figure}

\begin{figure}[htbp]
  \begin{center}
\includegraphics[width=.5\textwidth]{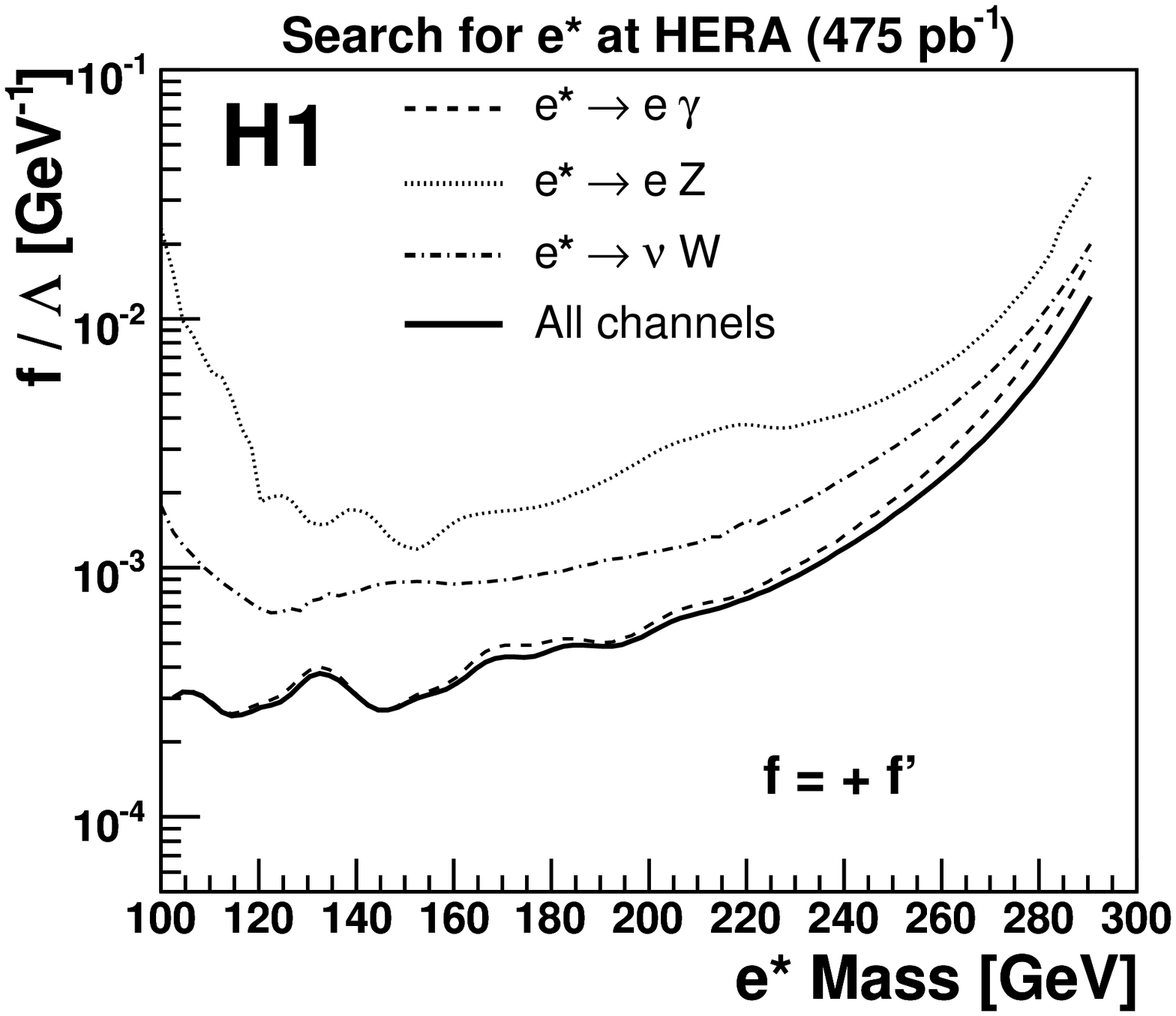}\put(-10,13){{\bf (a)}}
\includegraphics[width=.5\textwidth]{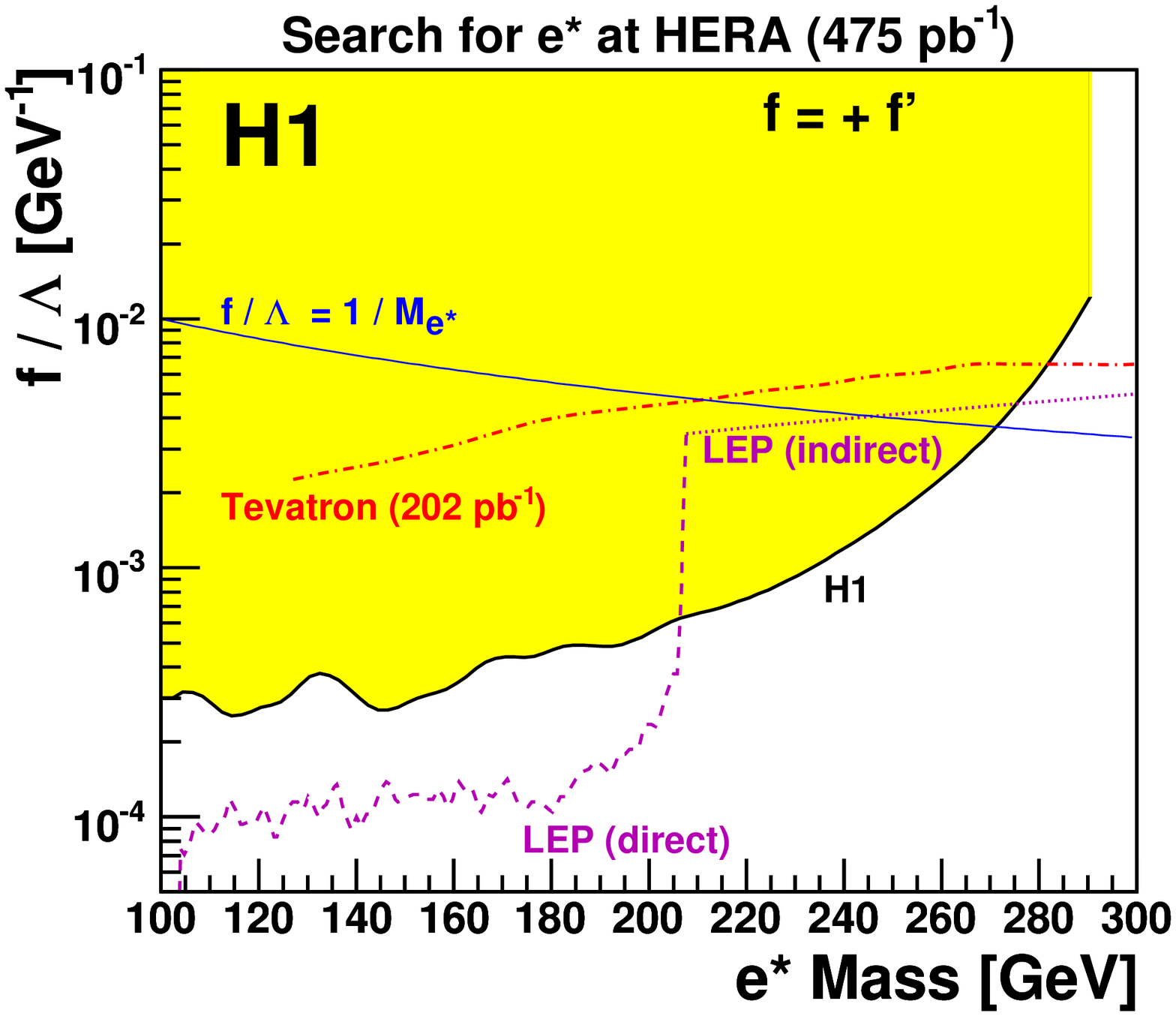}\put(-10,13){{\bf (b)}}
  \end{center}
  \caption{Exclusion limits  at $95$\% CL on the coupling $f/\Lambda$  as a function of the mass of the excited electron considering gauge mediated interactions only, with the assumption $f = +f'$.
The separate contributions of the different $e^*$ decay channels are presented in (a).  
Values of the couplings above the curves are excluded.
The excluded domain based on  all H1 $e^\pm p$ data is represented in (b) by the shaded area.
It is compared to the direct (dashed line) and indirect (dotted line) exclusion limits obtained at LEP by the OPAL Collaboration~\protect{\cite{Abbiendi:2002wf}} and by the DELPHI Collaboration~\protect{\cite{Abdallah:2004rc}}, respectively.
The result from the Tevatron obtained by the CDF experiment \protect{\cite{Acosta:2004ri}} is also shown (dashed-dotted line).
The curve $f/\Lambda = 1/M_{e^*}$ is indicated in (b).}
\label{fig:LimitCoupling}  
\end{figure}

Upper limits on the product of the $e^*$ production cross section and of the $e^*$ decay branching ratio are shown in figure~\ref{fig:LimitSBR}. 
The analysed decay channels of the $W$ and $Z$ gauge bosons are combined.
Considering pure gauge interactions, the resulting limit on $f/\Lambda$ after combination of all decay channels is displayed as a function of the $e^*$ mass in figure~\ref{fig:LimitCoupling}, for the conventional assumption $f = + f'$. 
The total fraction of all possible $e^*$ gauge decay channels covered in this analysis is $\sim88$\%.
The limit extends up to $e^*$ masses of $290$~GeV.
Considering the assumption $f/\Lambda = 1/M_{e^*}$ excited electrons with masses up to $272$~GeV are excluded. 
The relative contributions of the $e^*$ decay channels to the combined limit are shown in figure~\ref{fig:LimitCoupling}(a).
At low mass, the combined limit on $f/\Lambda$ is dominated by the $e^* {\rightarrow} e \gamma$ channel, while the $e^* \rightarrow \nu W$ channel starts to contribute to the limit for masses above $200$~GeV.
These new results extend the previously published limits by H1~\cite{Adloff:2002dy} and ZEUS~\cite{Chekanov:2001xk} by more than a factor of two in $f/\Lambda$.
Figure~\ref{fig:LimitCoupling}(b) shows direct and indirect limits on $e^*$ production obtained in $e^+ e^-$ collisions at LEP by the OPAL Collaboration~\protect{\cite{Abbiendi:2002wf}} and DELPHI Collaboration~\protect{\cite{Abdallah:2004rc}}, respectively.
The result of the most recent search for $e^*$ production within gauge mediated models obtained at the Tevatron by the CDF Collaboration is also indicated~\cite{Acosta:2004ri}. 
The limit from the present analysis extends at high mass beyond the kinematic reach of LEP searches and to lower $f/\Lambda$ values than are reached by Tevatron searches.

\begin{figure}[htbp]
  \begin{center}
\includegraphics[width=.5\textwidth]{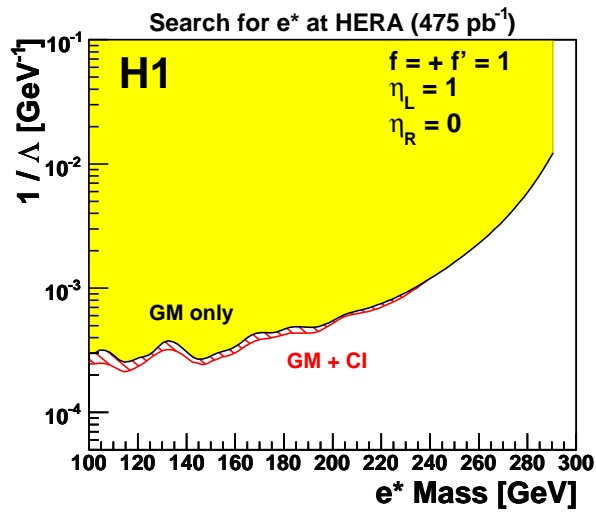}
  \end{center}
  \caption{Exclusion limits  at $95$\% CL on the inverse of the compositeness scale $1/\Lambda$  as a function of the mass of the excited electron.
The excluded domain obtained by considering $e^*$ production via gauge mediated interactions only and under the assumption  $f = +f' = 1$ is represented by the shaded area.
The hatched area corresponds to the additional domain excluded if gauge mediated and contact interactions are considered together for $e^*$ production.
Areas above the curves are excluded.}
\label{fig:Limit_CI}  
\end{figure}

If $e^*$ production is considered via gauge and contact interactions together, an upper limit on $1/\Lambda$ is also obtained, under the assumption $f = f' = 1$. 
Possible $e^*$ decays by either gauge or contact interactions are taken into account and the efficiency of the analysis to $e^*$ CI decays is conservatively assumed to be zero.
The limit on $1/\Lambda$ as a function of the $e^*$ mass is displayed in figure~\ref{fig:Limit_CI}.
For $e^*$ masses below $250$~GeV, the additional contribution of CI to $e^*$ production changes the limit on $\Lambda$ by a factor of  $1.15$ to $1.2$.
A limit on $\Lambda$ as a function of the $e^*$ mass is also obtained at the Tevatron by considering single $e^*$ production via contact interactions only, followed by its gauge decay into an electron and a photon~\cite{Abazov:2008hw}.

\section{Conclusion}

Using the full $e^{\pm}p$ data sample collected by the H1 experiment at HERA with an integrated luminosity of $475$~pb$^{-1}$ a search for the production of excited electrons is performed. 
The excited electron decay channels ${e}^{*} {\rightarrow} {e}{\gamma}$,  ${e}^{*} {\rightarrow} {e}{Z}$ and ${e}^{*} {\rightarrow} {\nu}{W}$ with subsequent hadronic or leptonic decays of the $W$ and $Z$ bosons are considered and no indication of a signal is found.
New limits on the production cross section of excited electrons are obtained. 
Within gauge mediated models, an upper limit on the coupling $f/\Lambda$ as a function of the excited electron mass is established for the specific relation $f = +f'$ between the couplings.
Assuming $f = + f'$ and $f/\Lambda=1/M_{e^*}$ excited electrons with a mass lower than $272$~GeV are excluded at $95$\% confidence level.
For the first time in $ep$ collisions, gauge and four-fermion contact interactions are also considered together for $e^*$ production and decays. In this scenario and assuming the same $\Lambda$ parameter in contact and gauge interactions as well as $f = + f' = 1$, $\eta_L = 1$ and $\eta_R =0$, the limit on $1/\Lambda$ improves only slightly, demonstrating that the gauge interaction mechanism is dominant for excited electron processes at HERA.
The results presented in this paper extend previously excluded domain at HERA, LEP or Tevatron.

\section*{Acknowledgements}

We are grateful to the HERA machine group whose outstanding
efforts have made this experiment possible. 
We thank the engineers and technicians for their work in constructing 
and maintaining the H1 detector, our funding agencies for financial 
support, the DESY technical staff for continual assistance and the 
DESY directorate for the hospitality which they extend to the non DESY 
members of the collaboration.
We also wish to thank M. Spira for many useful discussions and for providing the cross section calculation for excited electron production including contact interactions in $ep$ collisions.


\clearpage

\begin{table}[]
\begin{center}
\begin{tabular}{l c c c}
\multicolumn{4}{c}{{\bf Search for \begin{boldmath}$e^*$\end{boldmath} at HERA (\begin{boldmath}$475$\end{boldmath} pb\begin{boldmath}$^{-1}$\end{boldmath})}}\\
\hline
Channel & Data & SM & Signal Efficiency [\%]\\
\hline
${e}^{*} {\rightarrow} e \gamma$ (ela.) & $42$ & $48 \pm 4$ & $60$--$70$\\

${e}^{*} {\rightarrow} e \gamma$ (inel.) & $65$ & $65 \pm 8$ & $60$--$70$\\

${e}^{*} {\rightarrow} \nu W {\rightarrow} \nu q\bar{q}$ & $129$ & $133 \pm 32$ & $20$--$55$\\

\hspace{-0.2cm}\begin{tabular}{l} ${e}^{*} {\rightarrow} \nu W {\rightarrow} \nu e \nu$ \\ ${e}^{*} {\rightarrow} e Z {\rightarrow} e \nu \nu $ \end{tabular} &  $4$ & $4.5 \pm 0.7$ & \begin{tabular}{l} $60$ \\ $35$ \end{tabular}\\

${e}^{*} {\rightarrow} e Z {\rightarrow} e q\bar{q}$ & $286$ & $277 \pm 62$ & $20$--$55$ \\

 ${e}^{*} {\rightarrow} e Z {\rightarrow} eee$ & $0$ & $0.72 \pm 0.06$ & $60$ \\
 ${e}^{*} {\rightarrow} e Z {\rightarrow} e\mu\mu$ & $0$ & $0.52 \pm 0.05$ & $40$--$15$ \\
\hline
\end{tabular}
\end{center}

\caption{
Observed and predicted event yields for the studied $e^*$ decay channels.
  The analysed data sample corresponds to an integrated luminosity of $475$~pb$^{-1}$.
  The errors on the SM predictions include model and experimental systematic errors added in quadrature.
  Typical selection efficiencies for $e^*$  masses ranging from $120$ to $260$~GeV are also indicated.
}
\label{tab:estaryields}
\end{table}

\clearpage

\end{document}